\documentclass[aps,prb,twocolumn,superscriptaddress]{revtex4-1}
\usepackage{amssymb}
\usepackage{amsmath}
\usepackage{graphicx}
\usepackage{epsfig}

\begin{document}

\title{Probing quantum phase transition in a staggered Bosonic Kitaev chain
via layer-resolved localization-delocalization transition}
\author{R. Wang}
\author{X. Z. Zhang}
\email{zhangxz@tjnu.deu.cn}
\affiliation{College of Physics and Materials Science, Tianjin Normal University, Tianjin
300387, China}

\begin{abstract}
The bosonic statistics, which allow for macroscopic multi-occupancy of
single-particle states, pose significant challenges for analyzing quantum
phase transitions in interacting bosonic systems, both analytically and
numerically. In this work, we systematically investigate the non-Hermitian
Bloch core matrix of a Hermitian staggered bosonic Kitaev chain, formulated
within the Nambu framework. We derive explicit analytic conditions for the
emergence of exceptional points (EPs) in the $4\times 4$ Bloch core matrix,
with each EP marking the onset of complex-conjugate eigenvalue pairs. By
mapping the full many-body Hamiltonian onto an effective tight-binding
network in Fock-space and introducing layer-resolved inverse participation
ratio, we demonstrate that these EPs coincide precisely with sharp
localization--delocalization transitions of collective eigenstates.
Comprehensive numerical analyses across hopping amplitudes, pairing
strengths, and on-site potentials confirm that the EP of effective
Hamiltonian universally capture the global many-body phase boundaries. Our
results establish an analytically tractable, EP-based criterion for
detecting critical behavior in interacting bosonic lattices, with direct
relevance to photonic and cold-atom experimental platforms.
\end{abstract}

\maketitle


\section{Introduction}

The Kitaev model \cite{Kitaev}, originally formulated for fermionic systems,
has emerged as a paradigmatic platform for exploring topological phases of
matter, including topological superconductivity and Majorana zero modes. Its
appeal stems from the ability to capture nontrivial quantum phenomena within
a minimal and analytically tractable framework, establishing it as a
cornerstone of modern condensed matter physics. Motivated by the growing
interest in topological states, recent efforts have extended the Kitaev
paradigm to bosonic systems, leading to the bosonic Kitaev model \cite%
{McDonald1,Bardyn,Greiter}. This extension incorporates intrinsic bosonic
features, such as macroscopic state occupation and bosonic commutation
relations, while preserving the essential ingredients of pairing
interactions and spatially modulated hopping terms \cite{NJP1,AA1,Yoko,Ughre}%
. The bosonic Kitaev model provides a versatile platform to investigate
topological phase transitions and dissipative quantum dynamics in a setting
fundamentally distinct from its fermionic counterpart \cite{Bilit,Wilson1}.
The absence of the Pauli exclusion principle profoundly modifies the nature
of collective excitations, enabling macroscopic quantum coherence and
introducing rich dynamical responses to external perturbations, such as
staggered potentials and engineered dissipation. These distinctive features
make the model an ideal candidate for studying novel phases of matter,
particularly those driven by non-Hermitian physics, an area that has
attracted increasing attention in recent years \cite%
{McDonald1,NJP1,Flynn1,Pino,Slim,Wilson1}.

On the experimental front, arrays of nonlinear resonators and
superconducting circuits with engineered two-mode squeezing have recently
enabled the realization of bosonic pairing Hamiltonians with spatial
modulation. These platforms provide direct access to both the spectral
topology and damping dynamics of bosonic systems \cite%
{Sylvain,Qi,Wanjura,Hung,Wang,Sundar,Bilit,Wilson1}. From a theoretical
perspective, although the fermionic Kitaev chain admits an exact solution
via Jordan-Wigner transformation, its bosonic analogue resists such
analytical treatments. The absence of Pauli exclusion and the
non-commutativity of bosonic creation and annihilation operators result in
an infinite-dimensional many-body Hilbert space, rendering the quadratic
Hamiltonian non-diagonalizable via standard Bogoliubov transformations
except in certain limiting cases. These challenges motivate the development
of alternative approaches to understand the spectral and dynamical
properties of interacting bosonic systems.

To address these challenges, we develop a unified framework that maps the
dimerized bosonic Kitaev chain, which is subject to staggered on-site
potentials, onto effective single-particle tight-binding networks in
Fock-space. In two complementary limits, namely strong sublattice potential
imbalance and vanishing on-site potential, the non-Hermitian $4\times4$
momentum-space core matrix simplifies to either block-diagonal or purely
off-diagonal forms, respectively, allowing for the analytical determination
of exceptional points (EPs). Crossing these EPs coincides with the onset of
complex-conjugate eigenvalue pairs and, crucially, signals a sharp
transition from localized to delocalized many-body eigenstates, as diagnosed
by generalized layer-resolved inverse participation ratios (IPRs),
specifically the block IPR (BIPR$_1$) and block mean IPR (BMIPR$_1$) defined
in Fock-space layer coordinates.

Building on these limiting-case analyses, we then tackle the full parameter
regime by numerically computing the \textrm{BIPR}$_{2}$ and \textrm{BMIPR}$%
_{2}$ over the complete Fock-basis. We find that the hidden EP boundaries of
effective Hamiltonian continue to provide an accurate and practical
criterion for the bosonic localization--delocalization phase transition
across the entire parameter space. This EP-based criterion circumvents the
need for uncontrolled truncations of the bosonic Hilbert space, offering
instead a finite-dimensional diagnostic rooted in the analytic structure of
the non-Hermitian spectrum. Our approach thus lays the groundwork for
systematic explorations of critical behavior in driven interacting bosonic
lattices paving the way toward experimental tests in state-of-the-art
photonic and cold-atom simulators.

The remainder of this paper is organized as follows. In Sec.~\ref{Model
Hamiltonian}, we introduce the model Hamiltonian for a one-dimensional
bosonic Kitaev chain. Section~\ref{Reduced Hamiltonian and Exceptional
Points} analyzes the reduced Hamiltonian and identifies the EPs in two
limiting cases: strong sublattice potential imbalance and vanishing on-site
potential. In Sec.~\ref{Localization-Delocalization Transition}, we
demonstrate the localization-delocalization transition of many-body
eigenstates based on layer-resolved measures. Finally, Sec.~\ref{Summary}
summarizes our main findings and discusses their broader implications.

\section{Model Hamiltonian}

\label{Model Hamiltonian}

We consider a one-dimensional bosonic Kitaev chain with Hamiltonian $H=H_{%
\mathrm{T}}+H_{\mathrm{P}}$, where the kinetic term $H_{\mathrm{T}}$
incorporates staggered hopping and pairing interactions
\begin{eqnarray}
H_{\mathrm{T}} &=&\sum_{j=1}^{N}[it_{1}b_{2j-1}^{\dagger
}b_{2j}+it_{2}b_{2j}^{\dagger }b_{2j+1}+i\Delta _{1}b_{2j-1}^{\dagger
}b_{2j}^{\dagger }  \notag \\
&&+i\Delta _{2}b_{2j}^{\dagger }b_{2j+1}^{\dagger }+\mathrm{H.c.}].
\end{eqnarray}%
Here, $b_{l}^{\dagger }$ ($b_{l}$) creates (annihilates) a boson at site $l$%
, $t_{1,2}$ denote alternating hopping amplitudes, and $\Delta _{1,2}$
govern intra- and inter-dimer pairing strengths. The potential term $H_{%
\mathrm{P}}$ introduces on-site modulations
\begin{eqnarray}
H_{\mathrm{P}} &=&\sum_{j=1}^{N}[g_{1}(b_{2j-1}^{\dagger
}b_{2j-1}+b_{2j-1}b_{2j-1}^{\dagger })+g_{2}(b_{2j}^{\dagger }b_{2j}  \notag
\\
&&+b_{2j}b_{2j}^{\dagger })],
\end{eqnarray}%
where $g_{1,2}$ represent sublattice-dependent on-site potentials. All
parameters are real-valued, and periodic boundary conditions ($%
b_{j}=b_{2N+j} $) are imposed. The dimerized structure is explicit in the
sublattice operators $\alpha _{j}=b_{2j-1}$ and $\beta _{j}=b_{2j}$.
Exploiting translational invariance, we Fourier-transform the operators as
\begin{equation}
b_{l}=\frac{1}{\sqrt{N}}\sum_{k}e^{ikj}%
\begin{cases}
\alpha _{k}, & l=2j-1 \\
\beta _{k}, & l=2j%
\end{cases}%
,
\end{equation}%
with $k=2\pi m/N$ ($m=0,1,\dots ,N-1$). This decouples the Hamiltonian into
momentum sectors
\begin{equation}
H=\frac{1}{2}\sum_{-\pi \leqslant k<\pi }\left( H_{k}+H_{-k}\right) ,
\end{equation}%
where each $H_{k}$ is expressed in the Nambu basis $\Psi _{k}=\left( \beta
_{k},\alpha _{-k}^{\dagger },\alpha _{k},\beta _{-k}^{\dagger }\right) ^{T}$%
. Applying the unitary transformation $\Gamma =I_{2}\otimes \sigma _{z}$
with Pauli matrix $\sigma _{z}$, which preserves the canonical commutation
relations \cite{NJP1}. We cast $H_{k}=(\Gamma \Psi _{k})^{\dagger }h_{k}\Psi
_{k}$ with the non-Hermitian core matrix
\begin{equation}
h_{k}=%
\begin{pmatrix}
2g_{2} & i\Lambda _{k} & -iT_{k} & 0 \\
i\Lambda _{-k} & -2g_{1} & 0 & iT_{-k} \\
iT_{-k} & 0 & 2g_{1} & i\Lambda _{-k} \\
0 & -iT_{k} & i\Lambda _{k} & -2g_{2}%
\end{pmatrix}%
,  \label{hk0}
\end{equation}%
where $\Lambda _{k}=\Delta _{1}+\Delta _{2}e^{ik}$ and $%
T_{k}=t_{1}+t_{2}e^{ik}$ denote momentum-dependent pairing and hopping
functions, respectively. The eigenvalues of $h_{k}$ generally form four
energy bands, but this structure becomes modified in parameter regimes where
the non-Hermiticity dominates. The asymmetry in hopping ($t_{1}\neq t_{2}$)
and pairing ($\Delta _{1}\neq \Delta _{2}$) generates complex eigenvalues
that govern dissipative dynamics, fundamentally distinguishing this bosonic
system from its fermionic counterparts. On the other hand, we know that the
Hermitian system does not respect the complex system. Hence, the transition
from the real to complex is the key to understand the phase transition. The
transition point is referred to as the exceptional point (EP), at which the
eigestates coalesce. These spectral features critically influence the
stability of the system and topological classification, as elaborated in
Ref. \cite{NJP1}.
\begin{figure*}[tbp]
\centering
\includegraphics[bb=56 9 1793 1664, width=18 cm, clip]{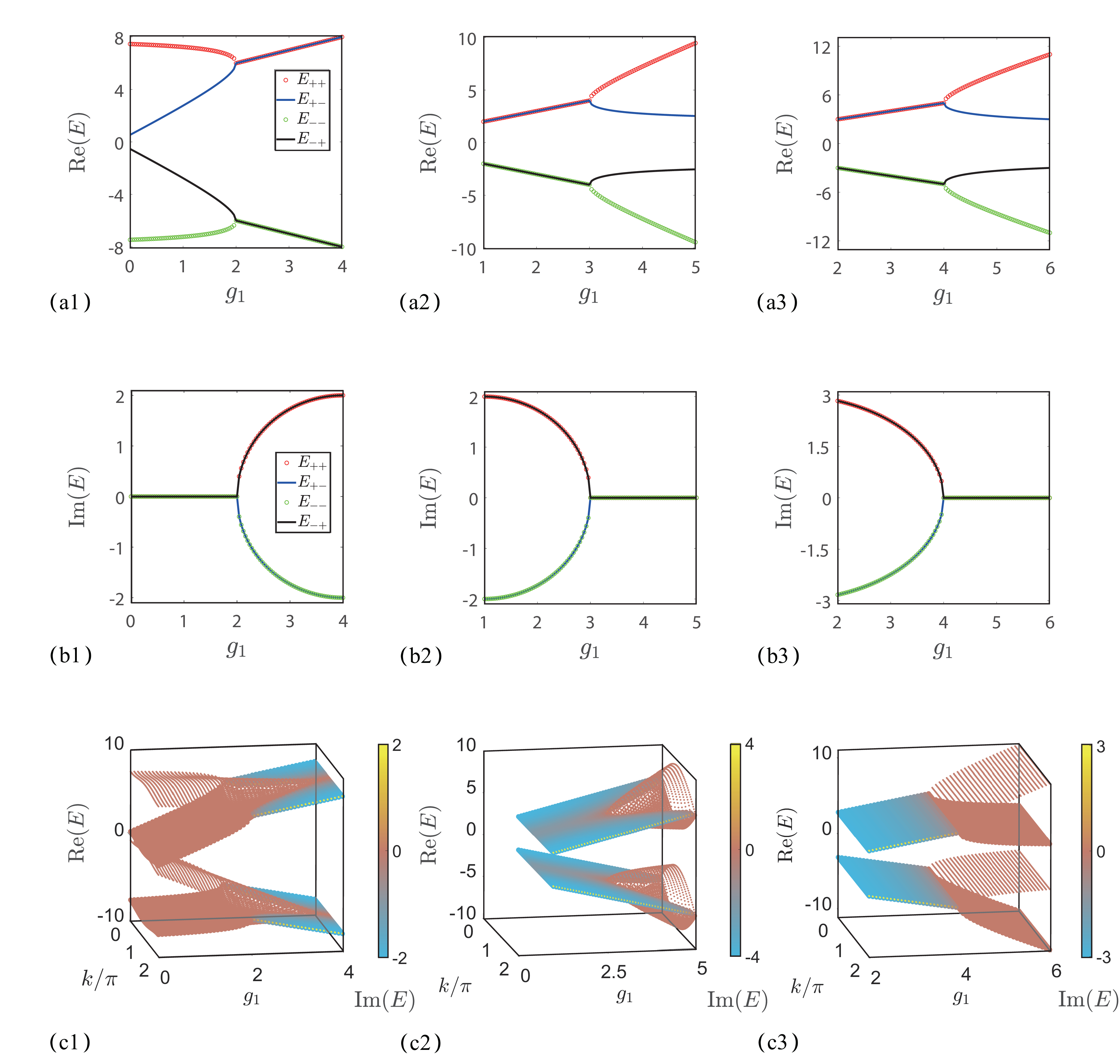}
\caption{Eigenenergy spectrum of the Hamiltonian $h_{k}$ in momentum space
under the strong sublattice potential limit, shown for three representative
sets of system parameters. Panels (a1-a3) and (b1-b3) display the real and
imaginary parts, respectively, of the eigenenergies of $h_{k}$ [Eq. (\protect
\ref{eq:block_diag})] as a function of the on-site potential $g_{1}$ at
specific momenta $k_{c}$. The system parameters are (a1,b1) $k_{c}=0$, $%
\Delta _{1}=\Delta _{2}=1$, $g_{2}=-4$; (a2,b2) $k_{c}=\protect\pi $, $%
\Delta _{1}=3\Delta _{2}=3$, $g_{2}=-1$; and (a3,b3) $k_{c}=\protect\pi $, $%
\Delta _{1}=3$, $\Delta _{2}=0$, $g_{2}=-1$, corresponding to the three
constraint equations in Eqs. (\protect\ref{eq:constr1})-(\protect\ref%
{eq:constr3}). Panels (c1-c3) depict the associated Riemann surface
structures of $h_{k}$ for these parameters. In all cases, the Hamiltonian $%
h_{k}$ hosts two second-order exceptional points (EP2) at $g_{1}=2$, $%
g_{1}=3 $, and $g_{1}=4$, respectively.}
\label{fig1}
\end{figure*}

\section{Reduced Hamiltonian and Exceptional Points}

\label{Reduced Hamiltonian and Exceptional Points}

We focus on the core matrix $h_{k}$, which encodes the essential physics of
the interacting bosonic system. Direct analytical diagonalization of this $%
4\times 4$ matrix presents significant challenges. To circumvent this, we
analyze two limiting cases that simplify the dynamics while preserving key
phenomena.

\subsection{Pairing-Dominated Regime}

Consider first the regime where a strong sublattice potential imbalance
dominates, satisfying
\begin{equation}
|g_{1}-g_{2}|\gg |T_{k}|,\quad g_{1}g_{2}<0.  \label{eq:reduce1}
\end{equation}%
Under this condition, the off-diagonal blocks of $h_{k}$ [Eq. (\ref{hk0})]
become negligible due to energy-scale separation, allowing the approximation
\begin{equation}
h_{k}\approx
\begin{pmatrix}
h_{k}^{(1)} & 0 \\
0 & h_{k}^{(2)}%
\end{pmatrix}%
,  \label{eq:block_diag}
\end{equation}%
with reduced $2\times 2$ sub-matrices
\begin{equation}
h_{k}^{(1)}=%
\begin{pmatrix}
2g_{2} & i\Lambda _{k} \\
i\Lambda _{-k} & -2g_{1}%
\end{pmatrix}%
,  \label{eq:hk1}
\end{equation}%
and
\begin{equation}
h_{k}^{(2)}=%
\begin{pmatrix}
2g_{1} & i\Lambda _{-k} \\
i\Lambda _{k} & -2g_{2}%
\end{pmatrix}%
.  \label{eq:hk2}
\end{equation}%
Diagonalizing Eq. (\ref{eq:block_diag}) yields the dispersion relation
\begin{equation}
E_{\rho \sigma }=\rho (g_{1}-g_{2})+\sigma \sqrt{(g_{1}+g_{2})^{2}-|\Lambda
_{k}|^{2}},  \label{eq:Eig1}
\end{equation}%
where $\rho ,\sigma =\pm 1$. The above dispersion relation predicts
band-touching EPs when
\begin{equation}
(g_{1}+g_{2})^{2}=|\Lambda _{k}|^{2}.  \label{eq:ep_condition}
\end{equation}%
Three distinct EP scenarios emerge:

(i) Brillouin zone center ($k_{c}=0$):
\begin{equation}
(g_{1}+g_{2})^{2}=(\Delta _{1}+\Delta _{2})^{2},  \label{eq:constr1}
\end{equation}

(ii) Brillouin zone edge ($k_{c}=\pi $):
\begin{equation}
(g_{1}+g_{2})^{2}=(\Delta _{1}-\Delta _{2})^{2},  \label{eq:constr2}
\end{equation}

(iii) Pairing localization ($\Delta _{2}=0$):
\begin{equation}
(g_{1}+g_{2})^{2}=\Delta _{1}^{2}.  \label{eq:constr3}
\end{equation}%
At these critical points, the Hamiltonian Eq. (\ref{eq:block_diag}) adopts a
Jordan block structure:
\begin{equation}
h_{k}^{\mathrm{JD}}=%
\begin{pmatrix}
g_{2}-g_{1} & 0 & 0 & 0 \\
1 & g_{2}-g_{1} & 0 & 0 \\
0 & 0 & g_{1}-g_{2} & 0 \\
0 & 0 & 1 & g_{1}-g_{2}%
\end{pmatrix}%
,  \label{eq:hJD1}
\end{equation}%
exhibiting two second-order EPs with critical energies $E_{c}=\pm
(g_{1}-g_{2})$. Crossing the EP boundary triggers a spectral transition:
eigenvalues evolve from complex-conjugate pairs to purely real (or vice
versa). For clarity, we plot the eigenenergy spectrum of the Hamiltonian $%
h_{k}$ in momentum space with three representative system parameters
[correspond to Eq. (\ref{eq:constr1}-\ref{eq:constr3})] and the Riemann
surface structures in Fig. \ref{fig1}.

This EP-mediated transition correlates with a fundamental change in
wavefunction structure from localized states in the gapped phase to
delocalized modes in the critical regime. The connection between EPs and
quantum phase transitions will be further explored in Section \ref%
{Localization-Delocalization Transition}.

\subsection{Absence of the On-site Potentials}

In the absence of on-site potential terms, i.e., when the system only
features competition between the hopping amplitudes $t_{1,2}$ and $\Delta
_{1,2}$ and pairing strengths, we impose the condition $g_{1}=g_{2}=0$.
Under this constraint, the non-Hermitian core matrix $h_{k}$ in Eq. (\ref%
{hk0}) can be recast as
\begin{equation}
h_{k}=\left(
\begin{array}{cccc}
0 & i\Lambda _{k} & -iT_{k} & 0 \\
i\Lambda _{-k} & 0 & 0 & iT_{-k} \\
iT_{-k} & 0 & 0 & i\Lambda _{-k} \\
0 & -iT_{k} & i\Lambda _{k} & 0%
\end{array}%
\right) ,  \label{hk2}
\end{equation}%
where $\Lambda _{k}=\Delta _{1}+\Delta _{2}e^{ik}$ and $%
T_{k}=t_{1}+t_{2}e^{ik}$ denote momentum-dependent pairing and hopping
functions, respectively. By direct diagonalization of Eq. (\ref{hk2}), we
obtain the quasi-particle dispersion relation
\begin{equation}
\Xi _{\rho \sigma }=\rho \sqrt{-|\Lambda _{k}|^{2}+|T_{k}|^{2}+\sigma \sqrt{%
(T_{-k}\Lambda _{k}-T_{k}\Lambda _{-k})^{2}}},  \label{Eig2}
\end{equation}%
where $\rho $, $\sigma =\pm 1$. To identify points where four energy bands
touch simultaneously, we impose the following constraint conditions
\begin{equation}
\left\{
\begin{array}{c}
-|\Lambda _{k}|^{2}+|T_{k}|^{2}=0 \\
T_{-k}\Lambda _{k}-T_{k}\Lambda _{-k}=0%
\end{array}%
\right. ,  \label{eq:constr4}
\end{equation}%
which, after simplification, yield
\begin{equation}
t_{1}t_{1}+t_{2}t_{2}+2t_{1}t_{2}\cos k-(\Delta _{1}\Delta _{1}+\Delta
_{2}\Delta _{2}+2\Delta _{1}\Delta _{2}\cos k)=0  \label{eq:constr51}
\end{equation}%
and
\begin{equation}
2i(t_{1}\Delta _{2}-t_{2}\Delta _{1})\sin k=0.  \label{eq:constr52}
\end{equation}%
For specific values of momentum, these equations reduce to simple forms:

(iv) At $k=0$, Eq. (\ref{eq:constr51}) and Eq. (\ref{eq:constr52}) reduce to
\begin{equation}
(t_{1}+t_{2})^{2}-(\Delta _{1}+\Delta _{2})^{2}=0,  \label{predi4}
\end{equation}

\begin{figure*}[tbp]
\centering
\includegraphics[bb=46 37 1764 1113, width=18 cm, clip]{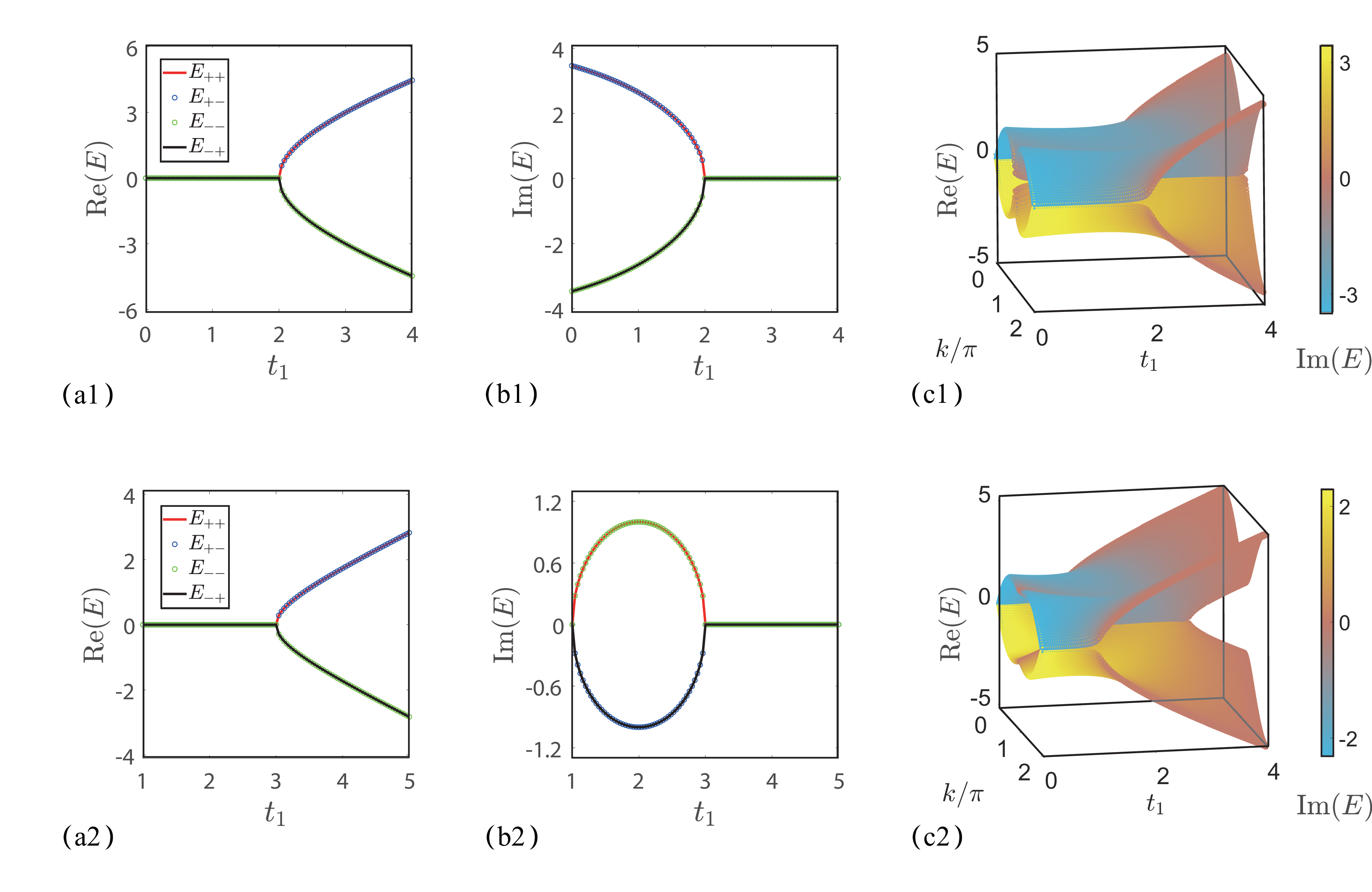}
\caption{Eigenenergy spectrum of the Hamiltonian $h_{k}$ [Eq. (\protect\ref%
{hk2})] in momentum space in the absence of on-site potential terms, shown
for two representative sets of system parameters. Panels (a1-a2) and (b1-b2)
display the real and imaginary parts, respectively, of the eigenenergies of $%
h_{k}$ as a function of the hopping parameter $t_{1}$ at specific momenta $%
k_{c}$. The system parameters are (a1,b1) $k_{c}=0$, $\Delta _{1}=3\Delta
_{2}=3$, $t_{2}=2$; and (a2,b2) $k_{c}=-\protect\pi $, $\Delta _{1}=2\Delta
_{2}=2$, $t_{2}=2$, corresponding to the two constraint conditions in Eqs. (%
\protect\ref{predi4})-(\protect\ref{predi5}). Panels (c1-c2) depict the
associated Riemann surface structures of $h_{k}$ for these parameters. In
both cases, the Hamiltonian $h_{k}$ hosts two second-order exceptional
points (EP2) at zero quasi-energy ($E_{c}=0$), occurring at $t_{1}=2$ and $%
t_{1}=3$, respectively.}
\label{fig2}
\end{figure*}

(v) At $k=-\pi $, Eq. (\ref{eq:constr51}) and Eq. (\ref{eq:constr52})
simplifies to
\begin{equation}
(t_{1}-t_{2})^{2}-(\Delta _{1}-\Delta _{2})^{2}=0.  \label{predi5}
\end{equation}%
Substituting the conditions of the above two cases into Eq. (\ref{hk2}),
direct calculations show that the non-Hermitian core matrix $h_{k}$ assumes
a Jordan block form with second-order exceptional points (EP2), given by
\begin{equation}
h_{k}^{JD}=\left(
\begin{array}{cccc}
0 & 0 & 0 & 0 \\
1 & 0 & 0 & 0 \\
0 & 0 & 0 & 0 \\
0 & 0 & 1 & 0%
\end{array}%
\right) .  \label{hJD2}
\end{equation}%
In Fig. \ref{fig2}, we numerically track the evolution of the band structure
as the system parameter $t_{1}$ varies. When the condition for band touching
given by Eqs. (\ref{eq:constr4})-(\ref{eq:constr52}) is satisfied, the
system undergoes a dramatic spectral transition: all eigenvalues abruptly
shift from real to complex-conjugate pairs. This behavior signals a quantum
phase transition characterized by the emergence of two second-order EP at
zero quasi-energy, i.e., $E_{c}=0$, under both parameter conditions of (iv)
and (v).

\section{Localization-Delocalization Transition}

\label{Localization-Delocalization Transition}

In this section, we extend our analysis of the bosonic Kitaev model to the
multi-particle space by introducing a Bardeen--Cooper--Schrieffer
(BCS)--like pairing basis, thereby mapping the problem onto an effective
single particle lattice. Our goal is to reveal hidden EP behavior in the
Hermitian system and to show that these EPs characterize the
localization--delocalization transition of the eigenstates.

Under the parameter constraint of Eq. (\ref{eq:reduce1}), namely that the
large band gap determined by on-site potential difference $|g_{1}-g_{2}|$
suppresses the transition $T_{\pm k}$, the non-Hermitian core matrix
acquires both boson number parity conservation
\begin{equation}
\lbrack \Pi _{1},H_{k}]=[\Pi _{2},H_{k}]=[\Pi _{\mathrm{total}},H_{k}]=0
\end{equation}%
and momentum conservation
\begin{equation}
\lbrack K_{1},H_{k}]=[K_{2},H_{k}]=[K_{\mathrm{total}},H_{k}]=0.
\end{equation}%
Here, the partial parity operators are
\begin{equation}
\left\{
\begin{array}{c}
\Pi _{1}=(-1)^{n_{\alpha ,-k}+n_{\beta ,k}} \\
\Pi _{2}=(-1)^{n_{\alpha ,k}+n_{\beta ,-k}}%
\end{array}%
\right. ,
\end{equation}%
and the total parity operator is
\begin{equation}
\Pi _{\mathrm{total}}=\Pi _{1}\Pi _{2}=(-1)^{n_{k}+n_{-k}}
\end{equation}%
with $n_{\pm k}=n_{\alpha ,\pm k}+n_{\beta ,\pm k}$, $n_{\alpha ,\pm
k}=\alpha _{\pm k}^{\dagger }\alpha _{\pm k}$, and $n_{\beta ,\pm k}=\beta
_{\pm k}^{\dagger }\beta _{\pm k}$. The corresponding partial momentum
operators and total partial momentum operator read
\begin{equation}
\left\{
\begin{array}{c}
K_{1}=-k(n_{\alpha ,-k}-n_{\beta ,k}) \\
K_{2}=k(n_{\alpha ,k}-n_{\beta ,-k})%
\end{array}%
\right. ,
\end{equation}%
and%
\begin{equation}
K_{\mathrm{total}}=K_{1}+K_{2}=k(n_{k}-n_{-k}).
\end{equation}%
We now restrict attention to the invariant subspace spanned by the following
Fock-basis
\begin{eqnarray}
|s,l_{s}\rangle &=&\frac{1}{\Omega _{1}}(\alpha _{-k}^{\dagger }\beta
_{k}^{\dagger })^{(s-l_{s})}(\alpha _{k}^{\dagger }\beta _{-k}^{\dagger
})^{(l_{s}-1)}  \notag \\
&&\times |0\rangle _{\alpha ,-k}|0\rangle _{\beta ,k}|0\rangle _{\alpha
,k}|0\rangle _{\beta ,-k},  \label{basis1}
\end{eqnarray}%
where $\Omega _{1}=(s-l_{s})!(l_{s}-1)!$ denotes the normalization
coefficient, $s$\ denotes the $s$th layer ($s=1$, $2$,..., $S$), $l_{s}$
labels the $l_{s}$th basis within $s$\ layer ($l_{s}=1$, $2$,..., $s$)$.$ $%
|0\rangle _{\alpha ,k}$ and$\ |0\rangle _{\beta ,k}$ are the vaccum state of
the bosonic operators $\alpha _{k}$ and $\beta _{k}$, respectively. One
readily verifies
\begin{equation}
\Pi _{1}|s,l_{s}\rangle =\Pi _{2}|s,l_{s}\rangle ,\Pi _{\mathrm{total}%
}|s,l_{s}\rangle =|s,l_{s}\rangle ,
\end{equation}%
and
\begin{equation}
K_{1}|s,l_{s}\rangle =K_{2}|s,l_{s}\rangle ,K_{\mathrm{total}%
}|s,l_{s}\rangle =0|s,l_{s}\rangle .
\end{equation}%
Grouping states by total boson number and examining the action of $H_{k}$ on
$\{|s,l_{s}\rangle \}$ one finds exact correspondence with a two-dimensional
tight-binding model featuring linearly growing nearest-neighbor hoppings and
on-site potentials. Explicitly, the effective Hamiltonian in this basis
takes the form
\begin{eqnarray}
H_{\mathrm{eq},1}^{k}
&=&\sum_{s=1}^{S}\sum_{l_{s}=1}^{s}\{[i(s-l_{s}+1)\Lambda
_{k}(|s+1,l_{s}\rangle \langle s,l_{s}|)  \notag \\
&&+il_{s}\Lambda _{-k}(|s+1,l_{s}+1\rangle \langle s,l_{s}|)+\text{\textrm{%
H.c.}}]  \notag \\
&&+2s(g_{1}+g_{2})(|s,l_{s}\rangle \langle s,l_{s}|)\}.  \label{Heff1}
\end{eqnarray}

\begin{figure*}[tbp]
\centering
\includegraphics[bb=21 30 819 434, width=15cm, clip]{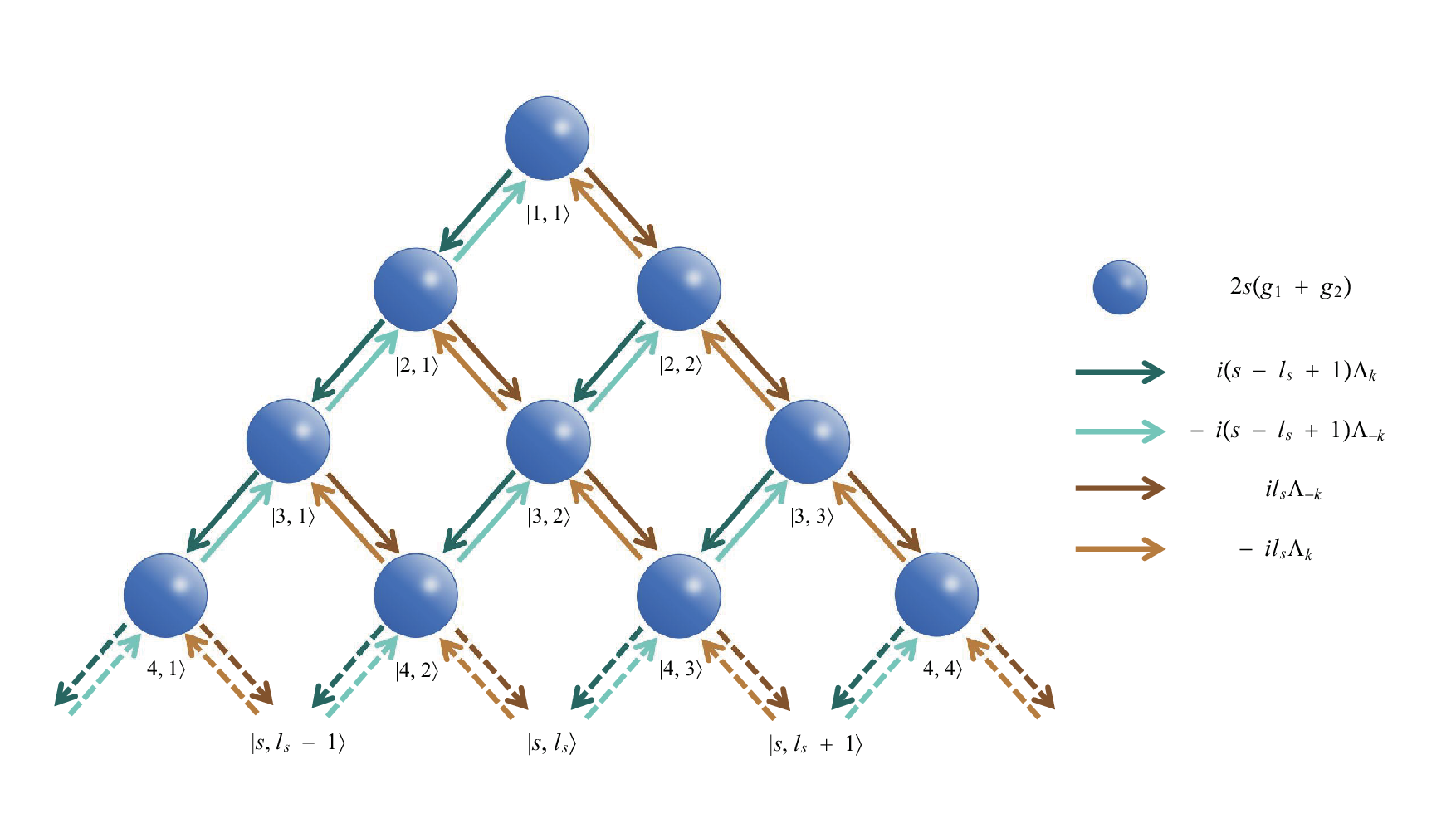}
\caption{Schematic illustration of the lattice structure corresponding to
the effective Hamiltonian in Eq. (\protect\ref{Heff1}). The lattice forms a
triangular structure composed of discrete sites with hopping and on-site
potential terms. Each lattice site is labeled as $|s,l_{s}\rangle $, where $%
s $ denotes the layer index and $l_{s}$ is the intra-layer site index within
layer $s$, comprising $s$ sites per layer. Blue solid spheres represent
sites with on-site potentials, and colored arrows indicate hopping between
neighboring sites across adjacent layers. It is noteworthy that there is no
intra-layer hopping, and the inter-layer hopping amplitudes increase with
the layer index $s$. All sites within the same layer share an identical
on-site potential value.}
\label{fig3}
\end{figure*}
Fig. \ref{fig3} depicts the lattice geometry associated with $H_{\mathrm{eq}%
,1}^{k}$, clearly illustrating how the bosonic many-body problem is mapped
onto a single-particle tight-binding network whose EPs govern the transition
between localized and delocalized eigenstates.

Inspired by Ref. \cite{HDK}, we propose that for the non-Hermitian core
matrix $h_{k}$ of Eq. (\ref{hk0}), the onset of EP behavior provides a sharp
criterion for the transition of eigenstates localization to delocalization:
For the real spectrum regime, all eigenvalues of $h_{k}$ are real and the
corresponding eigenstates are spatially localized. For the broken spectrum
regime, the complex-conjugate pairs emerge accompanied by the delocalization
of the eigenstates across the whole lattice.

To quantify localization in a single-layer system, one defines the inverse
participation ratio (\textrm{IPR}) of the $m$th single-particle eigenstate $%
|\varphi _{m}\rangle $ as
\begin{equation}
\text{\textrm{IPR}}(m)=\frac{\sum_{l}|\langle \varphi _{m}|l\rangle |^{4}}{%
(\sum_{l}|\langle \varphi _{m}|l\rangle |^{2})^{2}},
\end{equation}%
where $\{|l\rangle \}$ denotes the single-particle basis. For a $N$-site
lattice, finite-size scaling \textrm{IPR}$\propto N^{-\kappa }$ yields $%
\kappa =1$ for perfectly extended, $\kappa =0$ for totally localized states,
and $0<\kappa <1$ for intermediate cases. On the other hand, in systems that
traverse distinct quantum phases, the change in localization is not governed
by the behavior of a single eigenstate but by the collective response of
many---indeed, all---eigenstates. To capture this global
localization--delocalization transition as faithfully as possible, one
therefore introduces the mean inverse participation ratio (MIPR), defined as
the arithmetic average of the individual \textrm{IPR} over the full spectrum%
\begin{equation}
\text{\textrm{MIPR}}=\frac{\sum_{m=1}^{M}\text{\textrm{IPR}}(m)}{M},
\end{equation}%
where $M$ denotes the total number of the eigenstates.

For the specific effective Hamiltonian of Eq. (\ref{Heff1}), the Fock-space
\textquotedblleft layers\textquotedblright\ corresponding to different total
boson numbers have unequal dimensions. Consequently, the usual definitions
of the \textrm{IPR} and its \textrm{MIPR} must be changed to reflect this
layered structure. We therefore introduce the block inverse participation
ratio (\textrm{BIPR}) and its mean, \textrm{BMIPR}. For the $m$th eigenstate
$|\varphi _{m}\rangle $, we define the layer-resolved \textrm{IPR},
\begin{equation}
\mathrm{B}\text{\textrm{IPR}}_{1}(m)=\frac{\sum_{s=1}^{S}(|%
\sum_{l_{s}=1}^{s}\langle \varphi _{m}|s,l_{s}\rangle |^{4})}{%
[\sum_{s=1}^{S}(|\sum_{l_{s}=1}^{s}\langle \varphi _{m}|s,l_{s}\rangle
|^{2})]^{2}}.
\end{equation}%
The corresponding mean over all states is
\begin{equation}
\text{\textrm{BMIPR}}_{1}=\frac{\sum_{m=1}^{M_{1}}\mathrm{B}\text{\textrm{IPR%
}}_{1}(m)}{M_{1}},
\end{equation}%
where the total number of Fock-states satisfies
\begin{equation}
M_{1}=\sum_{s=1}^{S}s=\frac{S(S+1)}{2}.
\end{equation}%
By construction, \textrm{BIPR}$_{1}(m)\rightarrow 0$ if $|\varphi
_{m}\rangle $ is evenly distributed over all layers, and \textrm{BIPR}$%
_{1}(m)\rightarrow 1$ if it resides entirely within a single layer.

\begin{figure*}[tbp]
\centering
\includegraphics[bb=79 72 1750 534, width=18cm, clip]{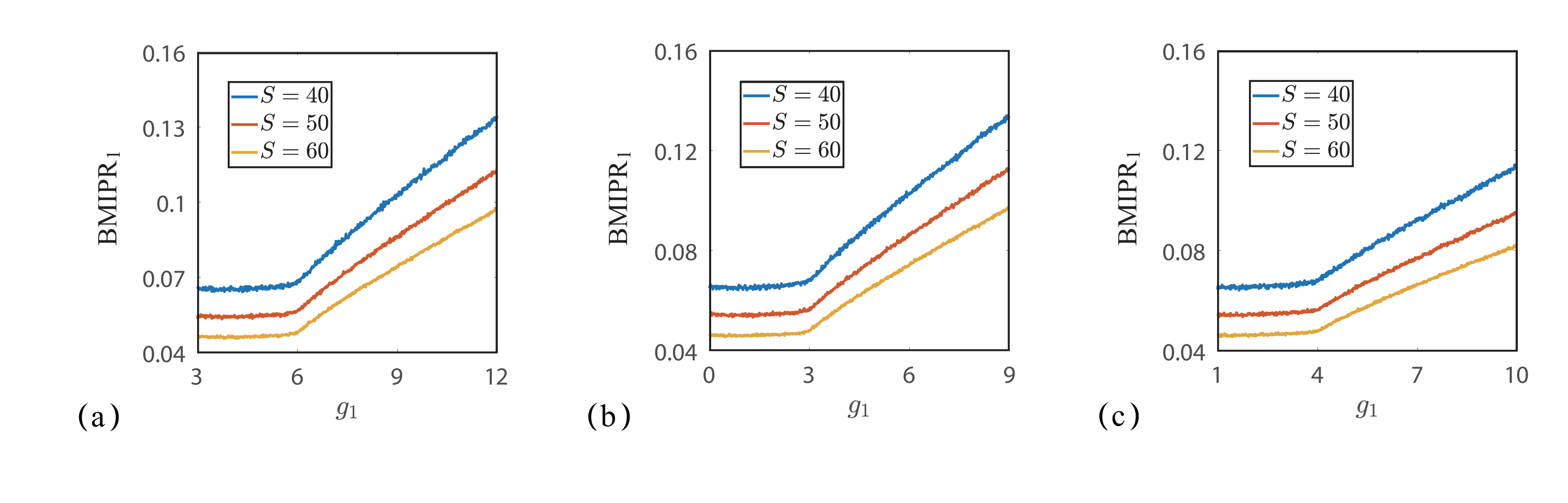}
\caption{Numerical results of the \textrm{BMIPR}$_{1}$ as a function of the
on-site potential parameter $g_{1}$ under three distinct sets of system
parameters. In all cases, a clear crossover point is observed at $g_{1}=6$, $%
g_{1}=3$, and $g_{1}=4$ in panels (a), (b), and (c), respectively. For
parameter values to the left of each crossover point, the system wave
functions exhibit extended-state characteristics, while to the right they
display localized-state behavior. These numerically determined critical
points agree excellently with the theoretical predictions from Eqs. (\protect
\ref{eq:constr1})-(\protect\ref{eq:constr3}). The system parameters are: (a)
$k_{c}=0$, $\Delta _{1}=\Delta _{2}=1$, $g_{2}=-4$; (b) $k_{c}=\protect\pi $%
, $\Delta _{1}=3\Delta _{2}=3$, $g_{2}=-1$; and (c) $k_{c}=\protect\pi $, $%
\Delta _{1}=3$, $\Delta _{2}=0$, $g_{2}=-1$. Blue, red, and yellow solid
lines correspond to system sizes $S=40$, $50$, and $60$, respectively.}
\label{fig4}
\end{figure*}

In Fig. \ref{fig4}, we present numerical simulations of \textrm{BMIPR}$_{1}$
as the system approaches its EP critical parameters for three representative
parameter sets: (I) $k_{c}=0$, $\Delta _{1}=\Delta _{2}$, $\Delta _{2}=1$, $%
g_{2}=-4$, (II) $k_{c}=\pi $, $\Delta _{1}=3\Delta _{2}$, $\Delta _{2}=1$, $%
g_{2}=-1$, and (III) $k_{c}=\pi $, $\Delta _{1}=3$, $\Delta _{2}=0$,$\
g_{1}=-1$. In each case, \textrm{BMIPR}$_{1}$ exhibits a sharp change in the
vicinity of the EP. The numerically extracted critical values $g_{1,c}=2$, $%
6 $ (case I), $g_{1,c}=-1$, $3$ (case II), and $\Delta _{1,c}=-2$, $4$ (case
III) are in excellent agreement with the theoretical predictions of Eqs. (%
\ref{eq:constr1})-(\ref{eq:constr3}), respectively.

In the regime where the on-site potential imbalance $|g_{1}-g_{2}|$ is no
longer large enough to suppress inter-band transitions induced by $T_{\pm k}$%
, the system dynamics depends nontrivially on the hopping amplitudes $%
t_{1,2} $, pairing strength $\Delta _{1,2}$, and on-site potentials $g_{1,2}$%
. In what follows, we focus on the Hamiltonian $H_{k}$ defined in the Eq. (%
\ref{hk0}) and show how to construct an effective single-particle lattice
model that captures the localization--delocalization quantum phase
transition of the general bosonic Kitaev system.

In this more general setting, only the total boson-parity and total momentum
remain conserved, i.e., $[\Pi _{\mathrm{total}},$ $H_{k}]=[K_{\mathrm{total}%
},$ $H_{k}]=0$. By contrast, the partial-parity and partial-momentum
operators cease to commute with $H_{k}$, i.e., $[\Pi _{i},H_{k}]\neq 0$, and
$[K_{i},H_{k}]\neq 0$ ($i=1,2$). Accordingly, the natural Fock-space basis
generalizes to three-index states
\begin{eqnarray}
|s,l_{s,1},l_{s,2}\rangle &=&\frac{1}{\Omega _{2}}(\alpha _{-k}^{\dagger
})^{(l_{s,1}-1)}(\beta _{k}^{\dagger })^{(l_{s,2}-1)}  \notag \\
&&\times (\alpha _{k}^{\dagger })^{(s-l_{s,2})}(\beta _{-k}^{\dagger
})^{(s-l_{s,1})}  \notag \\
&&\times |0\rangle _{\alpha ,-k}|0\rangle _{\beta ,k}|0\rangle _{\alpha
,k}|0\rangle _{\beta ,-k},  \label{basis2}
\end{eqnarray}%
where $\Omega _{2}=\sqrt{(l_{s,1}-1)!(l_{s,2}-1)!(s-l_{s,2})!(s-l_{s,1})!}$
denotes the normalization coefficient, $s$\ denotes the $s$th layer, $%
l_{s,1} $\ ($l_{s,2}$) labels the $l_{s,1}$th row ($l_{s,2}$th column) basis
within $s $\ layer ($l_{s,1(2)}=1,...s$). The\ above equation follows the
total boson parity and total momentum conservation.

By regrouping the Fock-space states of Eq. (\ref{basis2}) according to total
boson number and examining their structure, one can immediately map $H_{k}$
onto the effective single-particle tight-binding model on a layered lattice.
Hence, the effective Hamiltonian takes the form
\begin{eqnarray}
H_{eq,2}^{k} &=&\sum_{s=1}^{S}\sum_{l_{s,1},l_{s,2}=1}^{s}\{[(H_{\mathrm{%
intra,1}}+H_{\mathrm{intra,2}}+H_{\mathrm{inter,1}}  \notag \\
&&+H_{\mathrm{inter,2}})+\text{\textrm{H.c.}}]+H_{\mathrm{on-site}}\},
\label{Heff2}
\end{eqnarray}%
where
\begin{equation}
H_{\mathrm{intra,1}}=iT_{-k}\sqrt{l_{s,1}(s-l_{s,1})}(|s,l_{s,1}+1,l_{s,2}%
\rangle \langle s,l_{s,1},l_{s,2}|),  \label{H_intra_1}
\end{equation}%
\begin{equation}
H_{\mathrm{intra,2}}=-iT_{k}\sqrt{l_{s,2}(s-l_{s,2})}(|s,l_{s,1},l_{s,2}+1%
\rangle \langle s,l_{s,1},l_{s,2}|),  \label{H_intra_2}
\end{equation}%
\begin{equation}
H_{\mathrm{inter,1}}=i\Lambda _{k}\sqrt{l_{s,1}l_{s,2}}%
(|s+1,l_{s,1}+1,l_{s,2}+1\rangle \langle s,l_{s,1},l_{s,2}|),
\label{H_inter_1}
\end{equation}%
\begin{eqnarray}
H_{\mathrm{inter,2}} &=&i\Lambda _{-k}\sqrt{(s-l_{s,1}+1)(s-l_{s,2}+1)}
\notag \\
&&\times (|s+1,l_{s,1},l_{s,2}\rangle \langle s,l_{s,1},l_{s,2}|),
\label{H_inter_2}
\end{eqnarray}%
and
\begin{eqnarray}
H_{\mathrm{on-site}} &=&2[g_{1}(s+l_{s,1}-l_{s,2})+g_{2}(s-l_{s,1}+l_{s,2})]
\notag \\
&&\times (|s,l_{s,1},l_{s,2}\rangle \langle s,l_{s,1},l_{s,2}|).
\label{H_onsite}
\end{eqnarray}%
Accordingly, we generalize the general \textrm{BIPR} construction to the
second effective model by defining
\begin{equation}
\mathrm{B}\text{\textrm{IPR}}_{2}(m)=\frac{\sum_{s=1}^{S}(|%
\sum_{l_{s,1}=1}^{s}\sum_{l_{s,2}=1}^{s}\langle \varphi
_{m}|s,l_{s,1},l_{s,2}\rangle |^{4})}{[\sum_{s=1}^{S}(|\sum_{l_{s,1}=1}^{s}%
\sum_{l_{s,2}=1}^{s}\langle \varphi _{m}|s,l_{s,1},l_{s,2}\rangle |^{2})]^{2}%
}.
\end{equation}

\begin{figure*}[tbp]
\centering
\includegraphics[bb=90 4 1264 861, width=18cm, clip]{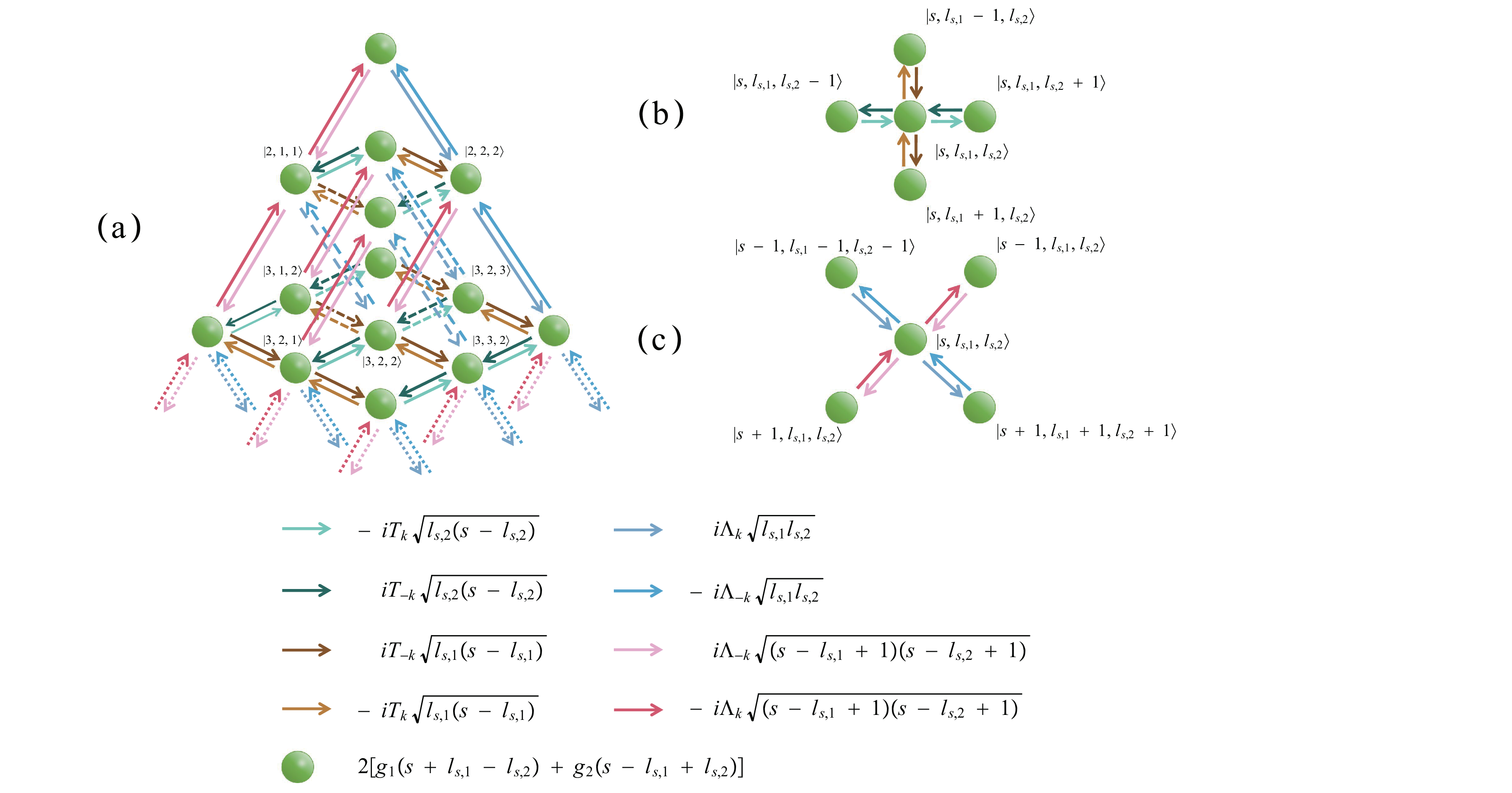}
\caption{Schematic illustration of the lattice structure corresponding to
the effective Hamiltonian in Eq. (\protect\ref{Heff2}). In panel (a), the
system forms a pyramidal lattice embedded in three-dimensional space,
incorporating intra-layer and inter-layer hopping terms along with on-site
potentials. Each lattice site is labeled as $|s,l_{s,1},l_{s,2}\rangle $,
where $s$ denotes the layer index and $(l_{s,1},l_{s,2})$ are the row and
column indices within layer $s$, comprising $s^{2}$ sites per layer. Green
solid spheres represent sites with on-site potentials, and colored arrows
indicate hopping processes within and between layers. Two distinct types of
inter-layer hopping are present: (i) a layer-dependent hopping whose
amplitude increases with $s$; and (ii) a layer-independent hopping. The
presence of layer-dependent inter-layer hopping leads to a rich landscape of
localization-delocalization phase boundaries in the system. Panels (b) and
(c) present the top view and side view of panel (a), respectively. Each site
in the structure has up to four intra-layer (inter-layer) nearest-neighbor
hopping terms.}
\label{fig5}
\end{figure*}
The corresponding mean over all states is
\begin{equation}
\mathrm{B}\text{\textrm{MIPR}}_{2}=\frac{\sum_{m=1}^{M_{2}}\mathrm{B}\text{%
\textrm{IPR}}_{2}(m)}{M_{2}},
\end{equation}%
where the total number of Fock-states satisfies
\begin{equation}
M_{2}=\sum_{s=1}^{S}s^{2}.
\end{equation}%
In Fig. \ref{fig5}, we present a schematic of the tight-binding lattice
corresponding to the effective Hamiltonian $H_{eq,2}^{k}$, where the $s$th
layer contains $s^{2}$ sites. In this network, there are three types of
interaction terms: (i) nearest-neighbor intra-layer hopping terms [see Eqs. (%
\ref{H_intra_1}-\ref{H_intra_2})]; (ii) inter-layer hopping terms connecting
sites in adjacent layers [see Eqs. (\ref{H_inter_1})-(\ref{H_inter_2})]; and
(iii) on-site potential terms acting on each lattice site [see Eq. (\ref%
{H_onsite})]. We emphasize that the inter-layer hopping terms exhibit two
distinct features. On one hand, in Hamiltonian $H_{\mathrm{inter,1}}$ [see
Eq. (\ref{H_inter_1})], the hopping amplitude depends only on the
intra-layer indices $l_{s,1}$ and $l_{s,2}$ of the sites involved, but is
independent of the layer index $s$. On the other hand, in Hamiltonian $H_{%
\mathrm{inter,2}}$ [see Eq. (\ref{H_inter_2})], the hopping amplitude
depends not only on the intra-layer site indices but also explicitly on the
layer index $s$.

\begin{figure*}[tbp]
\centering
\includegraphics[bb=82 68 1154 532, width=16cm, clip]{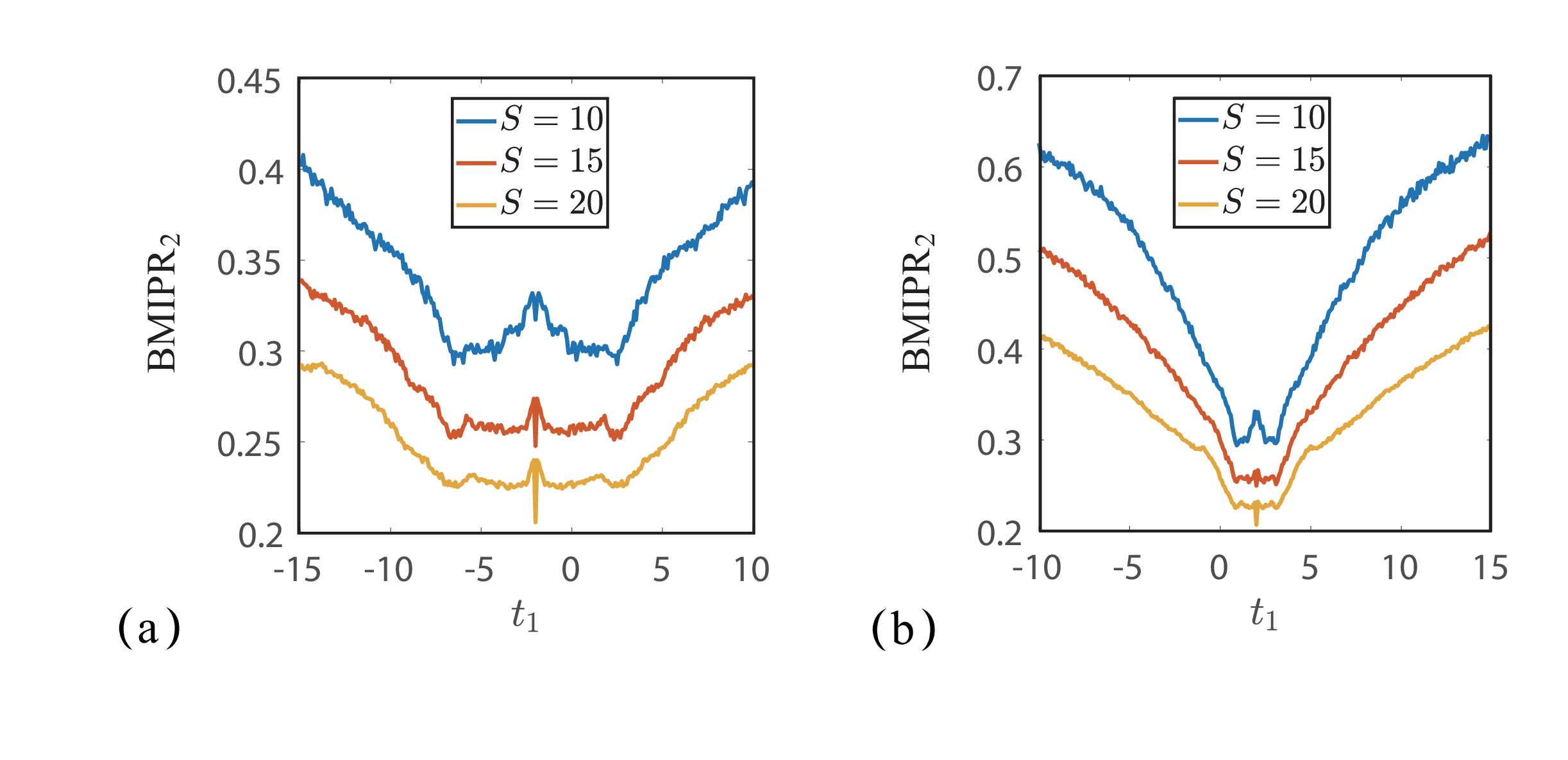}
\caption{Numerical results of the \textrm{BMIPR}$_{2}$ as a function of the
intra-layer nearest-neighbor hopping parameter $t_{1}$ for two
representative sets of system parameters. In panels (a) and (b), a clear
crossover point appears at $t_{1}=2$ and $t_{1}=3$, respectively. For
parameter values to the left of the crossover, the system's wavefunctions
exhibit extended behavior, whereas to the right of the crossover, they show
localized behavior. The numerical results agree well with the theoretical
predictions given in Eqs. (\protect\ref{predi4})-(\protect\ref{predi5}). The
system parameters are (a) $k_{c}=0$, $\Delta _{1}=3\Delta _{2}$, $\Delta
_{2}=1$, $t_{2}=2$; and (b) $k_{c}=-\protect\pi $, $\Delta _{1}=2\Delta _{2}$%
, $\Delta _{2}=1$, $t_{2}=2$. Blue, red, and yellow solid lines correspond
to system size parameters $S=10$, $15$, and $20$, respectively.}
\label{fig6}
\end{figure*}

Although we now possess both the analytic form of Eq. (\ref{Heff2}), a full
understanding of the eigenstate localization--delocalization transition
still relies on numerical diagonalization. Now we compute $\mathrm{B}$%
\textrm{MIPR}$_{2}$ across the phase boundary to identify how EP criticality
controls the global localization properties of the spectrum. In the special
limit $g_{1,2}=0$, we use the effective Hamiltonian of Eq. (\ref{Heff2}) to
compute $\mathrm{B}$\textrm{MIPR}$_{2}$ in the vicinity of the EP for the
two parameter regimes introduced earlier [cases (iv) and (v)]. Specifically,
we examine (VI) $k=0$, $\Delta _{1}=3\Delta _{2}$, $\Delta _{2}=1$, $t_{2}=2$%
, (V) $k=-\pi $, $\Delta _{1}=2\Delta _{2}$, $\Delta _{2}=1$, $t_{2}=2$.
Fig. \ref{fig6} displays the numerical evolution of $\mathrm{B}$\textrm{MIPR}%
$_{2}$ as these systems cross their EPs. We extract critical values $%
t_{1,c}=-6$, $2$, and $\Delta _{2,c}=1$, $3$ for cases (iv) and (v),
respectively. These numerically determined EP locations show excellent
agreement with the analytic predictions of Eqs. (\ref{predi4})-(\ref{predi5}%
), providing strong evidence that the emergence of EPs marks the transition
between localized and delocalized eigenstates. The progressively enhanced
inter-layer hopping, which increases with the layer index $s$, gives rise to
a richer set of localization--delocalization phase transitions. These
transitions are systematically characterized by layer-resolved \textrm{IPR}.

So far, we have explored two limits of the bosonic Kitaev model. First, by
neglecting the intra-layer hopping amplitudes $t_{1,2}$ within each
equal-boson-number subspace, we focused on the competition between the
inter-layer hoppings $\Delta _{1,2}$ and the effective on-site potentials $%
g_{1,2}$, and mapped out the resulting localization--delocalization phase
boundaries. Second, by setting $g_{1,2}=0$, and instead retaining the
intra-layer hoppings $t_{1,2}$, we examined how the interplay between $%
t_{1,2}$ and $\Delta _{1,2}$ alone governs the transition. In both
scenarios, we have shown that the hidden EPs of the non-Hermitian $h_{k}$
serve as the markers for the transition of the many-body eigenstates from
localized to delocalized in real space.

\begin{figure*}[tbp]
\centering
\includegraphics[bb=81 69 1308 533, width=18cm, clip]{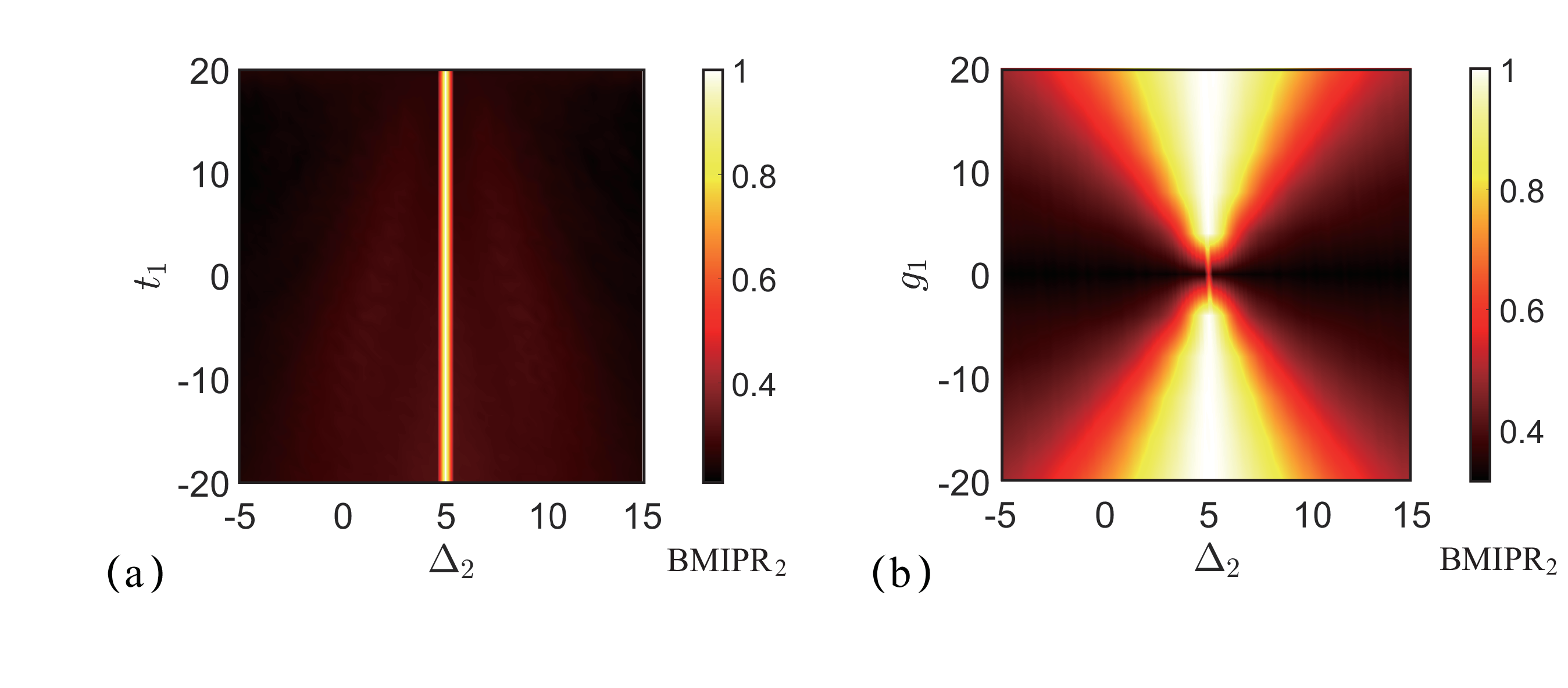}
\caption{Numerical results of the \textrm{BMIPR}$_{2}$ as a function of the
intra-layer hopping parameter $t_{1}$ and the on-site potential parameter $%
g_{1}$ for two representative sets of system parameters. In panel (a), the
system is tuned to the localization-delocalization phase boundary at $\Delta
_{2}=5$, consistent with the analytical condition derived from Eq. (\protect
\ref{eq:constr1}) under $t_{1}=t_{2}=0$. The phase boundary shows weak
sensitivity to variations in $t_{1}$, indicating robustness against
intra-layer perturbations. In panel (b), the system is tuned to the phase
boundary at $\Delta _{2}=5$, in agreement with the analytical prediction
from Eq. (\protect\ref{predi5}) under $g_{1}=g_{2}=0$. Here, the boundary
exhibits pronounced sensitivity to variations in $g_{1}$, yielding a richer
phase diagram. The system size is fixed at $S=20$. Other parameters are (a) $%
k_{c}=0$, $g_{1}=-g_{2}=10$, $t_{1}=t_{2}$, $\Delta _{1}=-5$; and (b) $k=-%
\protect\pi $, $t_{1}=t_{2}=10$, $g_{1}=g_{2}$, $\Delta _{1}=5$. Numerical
convergence tests confirm negligible finite-size effects and the robustness
of the observed phase boundaries. }
\label{fig7}
\end{figure*}

Building on the above results, we are particularly interested in how the
simultaneous competition among $t_{1,2}$, $\Delta _{1,2}$, and $g_{1,2}$
shapes the localization--delocalization phase boundary. Owing to the high
dimensionality of the parameter space, we restrict our attention to two
representative momentum sub-spces, $k_{c}=0$ and $k_{c}=-\pi $. In Fig \ref%
{fig7}, we present the numerically computed \textrm{BMIPR}$_{2}$ over the $%
\Delta _{2}-g_{1}-g_{2}$ parameter space, thereby describing the
localization--delocalization transition lines under full competition of $%
t_{1,2}$, $\Delta _{1,2}$, and $g_{1,2}$. For clarity, we choice two cases
with $k_{c}=0$, $g_{1}=-g_{2}=10$, $t_{1}=t_{2}$, $\Delta _{1}=-5$ [see Fig. %
\ref{fig7}(a)] and $k=-\pi $, $t_{1}=t_{2}=10$, $g_{1}=g_{2}$, $\Delta
_{1}=5 $ [see Fig. \ref{fig7}(b)], respectively. We note that the influence
of intra-layer hopping on the localization--delocalization phase boundary is
comparatively minor when intra-layer hopping, inter-layer hopping, and
on-site potential terms are all present. In contrast, inter-layer hopping
and on-site potential terms play a dominant role in determining the phase
boundary.

\section{Summary}

\label{Summary}

In this work, we introduced and systematically analyzed a dimerized bosonic
Kitaev chain subject to staggered on-site potentials, where asymmetric
hopping and pairing interactions combine to generate a non-Hermitian
quadratic Hamiltonian. By examining two analytically tractable limits,
namely strong sublattice potential imbalance and vanishing on-site
potential, we derived closed-form conditions for EPs in momentum space and
showed that each EP signals the emergence of complex conjugate eigenvalue
pairs. Mapping the full many-body problem onto effective tight-binding
networks in Fock-space layers, we demonstrated that these EPs coincide
precisely with sharp transitions between localized and delocalized
collective eigenstates, as diagnosed by layer-resolved inverse participation
ratios.

Beyond these limiting cases, numerical investigations across the full
parameter space, by varying hopping amplitudes, pairing strengths, and
on-site potentials, confirmed that the EP boundaries of the effective
Hamiltonian reliably predict the global many-body phase transition. This
EP-based diagnostic yields quantitatively accurate phase boundaries in
excellent agreement with analytic predictions. Our results establish a
robust, analytically grounded framework for identifying quantum phase
transitions in interacting bosonic lattices and offer clear experimental
signatures for observing EP-mediated critical phenomena in photonic and
cold-atom systems.

\acknowledgments We acknowledge the support of the National Natural Science
Foundation of China (Grants No. 12305026, 12275193, 11975166) and Science \&
Technology Development Fund of Tianjin Education Commission for Higher
Education(No. 2024KJ060).

\end{document}